\documentclass{osa-article}
\journal{OE}
\newcolumntype{M}[1]{>{\centering\arraybackslash}m{#1}}
\usepackage{braket}
\usepackage{diagbox}
\usepackage{multirow}

\newcommand{\eye}{\mathbb{I}} 

\begin{document}
\title{Accurate polarization preparation and measurement using twisted nematic liquid crystals}

\author{Martin Bielak,\authormark{1} Robert St\'arek,\authormark{1}, Vojt\v{e}ch Kr\v{c}marsk\'y,\authormark{1,2,3} Michal Mi\v{c}uda,\authormark{1} and Miroslav Je\v{z}ek\authormark{1,*}}

\address{\authormark{1}Department of Optics, Faculty of Science, Palacky University, 17. listopadu 12, 77146 Olomouc, Czechia}
\address{\authormark{2}Institut f\"ur Quantenoptik und Quanteninformation, \"Osterreichische Akademie der Wissenschaften, Technikerstr. 21A, 6020 Innsbruck, Austria}
\address{\authormark{3}Institut f\"ur Experimentalphysik, Universit\"at Innsbruck, Technikerstr. 25, 6020 Innsbruck, Austria}

\email{\authormark{*}jezek@optics.upol.cz}

\begin{abstract}
Generation of particular polarization states of light, encoding information in polarization degree of freedom, and efficient measurement of unknown polarization are the key tasks in optical metrology, optical communications, polarization-sensitive imaging, and photonic information processing. Liquid crystal devices have proved to be indispensable for these tasks, though their limited precision and the requirement of a custom design impose a limit of practical applicability. Here we report fast preparation and detection of polarization states with unprecedented accuracy using liquid-crystal cells extracted from common twisted nematic liquid-crystal displays. To verify the performance of the device we use it to prepare dozens of polarization states with average fidelity 0.999(1) and average angle deviation 0.5(3) deg. Using four-projection minimum tomography as well as six-projection Pauli measurement, we measure polarization states employing the reported device with the average fidelity of 0.999(1). Polarization measurement data are processed by the maximum likelihood method to reach a valid estimate of the polarization state. In addition to the application in classical polarimetry, we also employ the reported liquid-crystal device for full tomographic characterization of a three-mode Greenberger--Horne--Zeilinger entangled state produced by a photonic quantum processor.
\end{abstract}

\section{Introduction}
\label{sec:intro}

Direction, phase, and coherence of electromagnetic wave oscillation, i.e. its polarization state, represents an important feature determining the interaction of light and matter. Light reflection and scattering are fundamentally affected by incident polarization \cite{Huard1997}. Manipulating the state of polarization can significantly improve optical resolution \cite{Chen2019} and visualization of anisotropic structures in biomedical imaging \cite{VanEeckhout2019,He2021Sep,Zhanghao2019}. 
Polarization modulation enhances information capacity in optical communication \cite{Chen2016} and represents a building block of quantum communication and quantum information processing \cite{Flamini2018,Slussarenko2019}.
Precise generation and detection of the polarization state represent the crucial tasks in a vast number of applications.
Electrically tunable birefringent elements like liquid crystals, Pockels cells, and integrated electrooptical modulators are often employed to control polarization in such applications. 
Pockels based polarization modulators were successfully applied in optical switches \cite{Spagnolo2008} and loop-based photonic routing, where high transmittance is of paramount importance \cite{He2017,Takeda2019,Tiedau2019}. They offer high-speed operation and an acceptable extinction ratio, but their application is rather cumbersome due to their size and the necessity of high-voltage driving.
Integrated devices achieve even wider bandwidth enabling ultra-fast polarimetry \cite{Altepeter2011}. They possess significant losses and their calibration might be challenging, though, mainly due to unavoidable waveguide coupling.
Liquid crystals allow low-loss free-space polarization addressing using low-voltage control signals. Their operation is faster than mechanically manipulated birefringent elements, which has proven beneficial in polarimetry \cite{Bueno2000,DeMartino2003,Peinado2010,Peinado2011} and spectrometry \cite{Aharon2009,August2013}. Liquid crystal devices have also been successfully employed in polarization and phase modulation \cite{Zhuang1999,Moreno2007,Safrani2009,Peinado2014,Sciarrino2017,Sciarrino2018,Lohrmann2019} and switching \cite{Wang2008,Zhu2013}. Moreover, the absence of moving parts improves the robustness and lifespan of the device~\cite{Lohrmann2019, Perumangatt2021}.

In a vast majority of the polarimetric applications reported so far, nematic liquid crystals are used in custom devices acting as variable retarders \cite{Zhuang1999,Bueno2000,DeMartino2003,Safrani2009,Lohrmann2019}. Incompatibility of these devices with the common {\it twisted} nematic liquid crystal (TNLC) configuration widely utilized in display technology makes their broad application difficult.
Another complication stems from the fact, that additional wave plates or birefringent compensators are often utilized to change the overall polarization transformation.
Peinado et al. reported a polarimeter using a quarter-wave plate and a single tilted twisted nematic element in reflective geometry~\cite{Peinado2011}. The device was optimized to project an unknown polarization state to four polarization projections symmetrically placed on the Bloch sphere (Poincar\'e sphere), i.e. the vertices of a regular tetrahedron inscribed into the sphere. This minimum information measurement, also termed the minimum tomography \cite{Rehacek2004,Ling2006}, was shown to be efficient but prone to noise and measurement errors \cite{deBurgh2008,Ling2008,Bogdanov2010,Bogdanov2011,Koutny2016}.
Furthermore, a TNLC device applies a complex combination of rotation and retardation to incident polarization state, and the exact theoretical description of the device operation and its calibration remains a significant challenge \cite{Marquez2000,Yamauchi2005}.

Here we present a TNLC device based on a commercially available TNLC display with minimal modifications. The construction of the \emph{TNLC device} is described in Section~\ref{sec:construction}. In Section~\ref{sec:cellcalib}, we show that the theoretical model of a twisted nematic liquid crystals cell has a limited ability to predict the polarization changes. The novel calibration method, described in Section~\ref{sec:devcalib}, \emph{does not require} a theoretical model and yields unprecedented accuracy. We show how to calibrate the TNLC device to prepare an arbitrary polarization state and perform arbitrary polarization projection. The calibration stage is boosted by a genetic algorithm and other possible improvements are also discussed. In section~\ref{sec:preparation}, we test the TNLC device by preparing more than a hundred polarization states covering the Bloch sphere uniformly with an unprecedentedly high average fidelity of 0.999(1), including six eigenstates of Pauli operators and four states needed for the minimum tomography. Also, we show depolarized state preparation with degree of polarization 0.03(1). The presented device is ready to be used as a polarimeter, performing the polarization state tomography. In Section~\ref{sec:polarimter} we briefly review the polarization state tomography, and then we test the performance of the TNLC-based polarimeter. Furthermore, the TNLC device is employed in full quantum tomography of three-qubit Greenberger--Horne--Zeilinger entangled state produced by photonic quantum Toffoli gate.


\section{Construction of TNLC device}
\label{sec:construction}
\begin{figure}[ht!]
	\begin{center}
		\makebox[\linewidth]{
			\includegraphics[height=0.16\textheight]{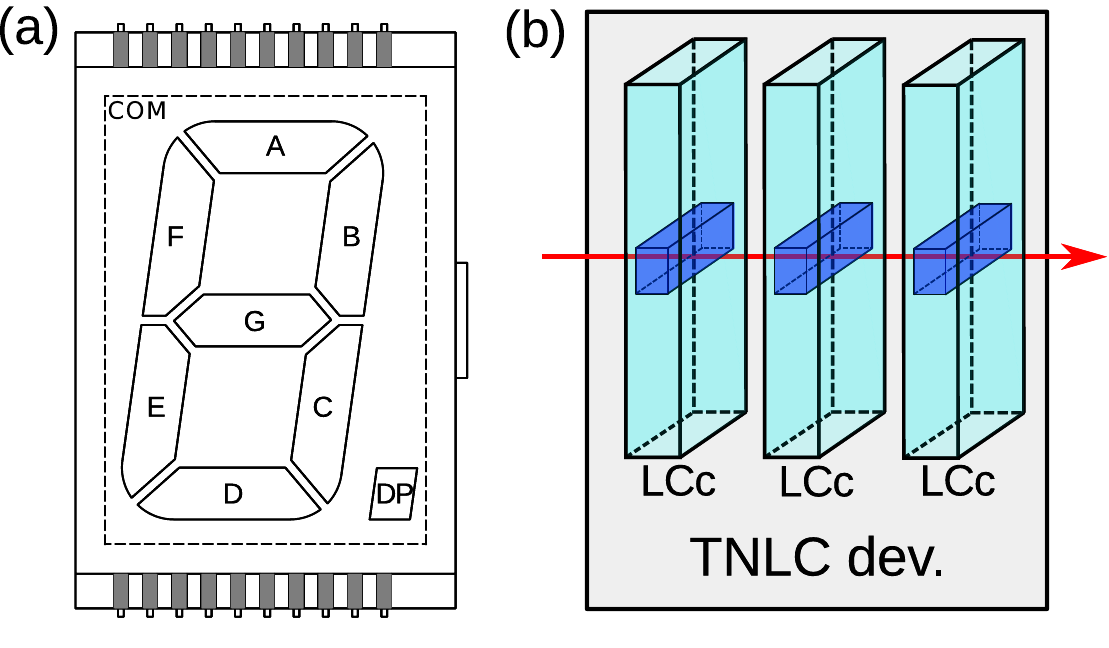}
			\includegraphics[height=0.16\textheight]{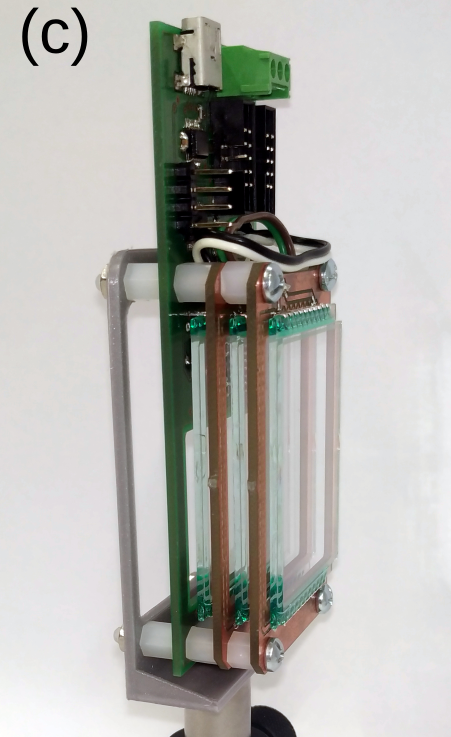}
		}
		\caption {			
			(a) Seven segment liquid crystal display layout; elements of the digit are labeled from A to G, the decimal point as DP, and the common back electrode as COM.
			(b) Scheme of the presented TNLC device consisting of three TNLC cells (LCc); dark blue blocks show the active segments and the red arrow shows the optical beam.
			(c) Photo of the TNLC device mounted on printed circuit board containing the electronic driver and communication interface.
		}
		\label{Fig. TN LCd}
	\end{center}
\end{figure}

The presented polarimetric device consists of three TNLC cells extracted from commercially available displays. We tested several commercially available displays, namely TN reflective display \emph{Lumex~LCD-S101D14TR}, super-TN (STN) reflective display \emph{RS~Pro~5080PHR}, transmissive TNLC display \emph{OCZ~Vrchlabi~2027}. We also tested transmissive STN pixel display \emph{OCZ~Vrchlabi~2001} and reflective STN pixel display \emph{DOG~XL}. Due to their limited fill factor, the transmitted polarization state is spatially modulated and effectively depolarized. We therefore find them inappropriate for the presented application. We selected display \emph{Lumex LCD-S101D14TR}, which has 7 segments (cells) and dimensions 33$\times$50~mm, see Fig.~\ref{Fig. TN LCd}(a). This particular model was chosen as a trade-off between the size of the segment and the overall compactness of the whole device. The TNLC reflective display cannot be directly used in transmission geometry or even to perform a unitary transformation. Therefore, we first removed reflective layers, protective films, polarizing sheets, and also other auxiliary layers from the displays. Using a micrometer screw gauge, we measured the thickness of the LC layer to be $6\pm2 \mu\mathrm{m}$. The number in parenthesis represents one standard deviation at the least significant digit.

We mounted the TNLC glass cells on printed circuit boards and stacked them, as shown in Fig.~\ref{Fig. TN LCd}(b,c). The main circuit board contains a driver consisting of a microcontroller, 16-bit digital-to-analog converter, and a voltage amplifier \cite{GitHub}. The microcontroller receives serial commands from a computer and controls the digital-to-analog converter. The produced control voltage signal is a symmetric square wave with frequency 1~kHz and 50$\%$ duty cycle. The square wave amplitude spans from 0 to 10 Vpp and effectively controls the action of the liquid crystals. For brevity, we refer to the control voltage amplitude simply as the \emph{voltage}. The voltage is applied between the central segment electrode, denoted G in Fig.~\ref{Fig. TN LCd}(a), and the common back electrode (COM) of the TNLC cell. The central segment has a 15$\times$4.5~mm approximately rectangular clear aperture, which is perfectly sufficient for collimated optical beams with mode field diameter up to 2.5 mm. The device can address the polarization state of several parallel optical beams arranged in a line or a matrix in complex quantum information processing circuits \cite{Lanyon2008,Micuda2013,Xiao2017}.

The transmittance of the single cell is 84\% and the transmittance of the whole device is 59\% at wavelength 810~nm. Limited transmittance is not an obstacle for classical polarimetry. At the single-photon level, the presented transmittance is also sufficient, as we demonstrate in Section~\ref{sec:polarimter}. Antireflective coating TNLC cells could increase the transmittance of a single cell up to $90\%$, increasing the overall transmittance to $73\%$. Only in loss-sensitive scenarios, for example measurements involving squeezed light, the limited transmittance becomes an obstacle.

\section{Characterization of a single TNLC cell}
\label{sec:cellcalib}

Let us start with a brief review of the Dirac notation, used here to describe the polarization, and connect it to the traditional description of polarized light. We describe pure polarization states with a \emph{ket state}, denoted with symbol $\ket{\cdot}$, which is equivalent to the normalized Jones vector. It holds that $\ket{J} = \frac{\mathbf{J}}{||\mathbf{J}||}$, where $\mathbf{J} = J_{x}\mathbf{x} + J_{x}\mathbf{y}$ is the Jones vector and $\mathbf{x}$ and $\mathbf{y}$ are orthonormal vectors, both normal to the direction of light propagation. The choice of basis is arbitrary, and the vector can be written as a superposition of any two orthonormal basis states. We choose basis states $\ket{H}$ and $\ket{V}$ corresponding to horizontally and vertically polarized light to keep the close correspondence between Jones vector and ket states. Explicitly, the correspondence is $\ket{J} = \left(J_{x}\ket{H} + J_{y}\ket{V}\right)\left(|J_{x}|^2 + |J_{y}|^2\right)^{-1/2}$.

Density matrix $\rho = \sum_{\lambda} p_\lambda |J_\lambda\rangle\langle J_\lambda|$ composed as an incoherent mixture of pure states $\ket{J_{\lambda}}$ with a probability distribution $p_\lambda$ also describes polarization of light, but in comparison to ket-vectors, it can also describe partially polarized light. Term $|J_\lambda\rangle\langle J_\lambda|$ is the outer product of $\ket{J}$ and its Hermitian conjugate, a $2\times2$ complex trace-normalized Hermitian matrix. Any polarization density matrix can be expanded using the identity $\eye$ and the Pauli matrices, $\sigma_1,\,\sigma_2,\,\sigma_3$, $\rho = (\eye + \sum_{j=1}^{3} u_j \sigma_j )/2$, where ${\textbf u}=(u_1,u_2,u_3)$ is the \emph{Bloch vector}. Compared to the Stokes vector, the Bloch vector does not contain information about the optical intensity, and its size is upper-bounded to one. 
We represent the polarization states geometrically using the \emph{Bloch sphere}. A Bloch vector contains Cartesian coordinates of a point within the sphere. A point on the surface of the Bloch sphere, described with longitude $\theta$ and latitude $\phi$, corresponds to a pure state $\ket{J} = \cos(\frac\theta2)\ket{H} + \exp(i\phi)\sin(\frac\theta2)\ket{V}$. In this representation, left- and right-handed circular polarizations lie on the south and north pole of the sphere, respectively. All linear polarizations lie on the equator of the sphere.
A point inside the sphere describes a partially polarized state. We quantify its \emph{purity} as $P = \mathrm{Tr}(\rho^2) = \frac{1}{2}\left(1 + ||\mathbf{u}||^2\right)$. The length of Bloch vector, $||\mathbf{u}|| = \sqrt{\sum\limits_{j=1}^{3}u_j^2}$, in the expression for purity is the \emph{degree of polarization} $\mathrm{DoP} = \sqrt{1-4\cdot \mathrm{det}(\rho)}$. \emph{Fidelity} $F = [\mathrm{Tr}\sqrt{\sqrt{\rho_u}\rho_v\sqrt{\rho_u}}]^2$ of density matrices $\rho_u$ and $\rho_v$ quantifies their similarity. It can be expressed in terms of the corresponding Bloch vectors, $\mathbf{u}$ and $\mathbf{v}$, as $F = \frac{1}{2}\left( 1 + \mathbf{u}\cdot\mathbf{v} + \sqrt{(1-||\mathbf{u}||)(1-||\mathbf{v}||)} \right)$. The fidelity is connected to the angular deviation of the Bloch vectors $\vartheta = \frac{1}{2}\arccos\left( \frac{\mathbf{u}\cdot\mathbf{v}}{||\mathbf{u}||\,||\mathbf{v}||} \right)$ via the dot-product term. 

The polarization transformation introduced by a device is described with matrix $M$ acting on a density matrix of the input state, $\rho_{\rm out} = M \rho_{\rm in} M^{\dagger}$, where $\dagger$ stands for Hermitian conjugation. The transformation matrix $M$ of an ideal TNLC cell can be derived using a sequence of thin wave plates, each introducing a small phase delay and oriented at linearly increasing azimuth angle \cite{YarivYeh1984}. The transformation reads
\begin{eqnarray}
M_{\text{TNLC}}&=&
R(\alpha)
R(\varphi)
\begin{pmatrix}
\cos{\chi}+i\dfrac{\delta}{\chi}\sin{\chi}&-\dfrac{\varphi}{\chi}\sin{\chi}\\
\dfrac{\varphi}{\chi}\sin{\chi}&\cos{\chi}-i\dfrac{\delta}{\chi}\sin{\chi}
\end{pmatrix}
R(\alpha)^{\dagger},\label{eq: TNLC matrix}\\
R(\varphi) &=& \begin{pmatrix}
\cos{\varphi}&\sin{\varphi}\\
-\sin{\varphi}&\cos{\varphi}
\end{pmatrix},
\end{eqnarray}
where $\chi=\sqrt{\varphi^2+\delta^2}$, $\varphi$ is the twist angle, $\alpha$ is the rotation angle of the cell, and $\delta$ represents the phase delay dependent on the applied voltage $V$.
The phase delay (\emph{retardance}) is almost constant up to a threshold voltage and decreases for larger voltages. The retardance is usually approximated using an $\arctan{}$ function \cite{Davis1998}. We propose to model the retardance more precisely using a logistic function,
\begin{eqnarray}
\delta=A+\dfrac{1}{B+\exp{(C-D V^E)}}, \label{eq: Logistic}
\end{eqnarray} 
where $A$, $B$, $C$, $D$, and $E$ are constants and $V$ is the voltage applied on the TNLC cell. 

\begin{figure}[ht!]
	\begin{center}
		\makebox[\linewidth]{
			\includegraphics[width=0.5\textwidth] {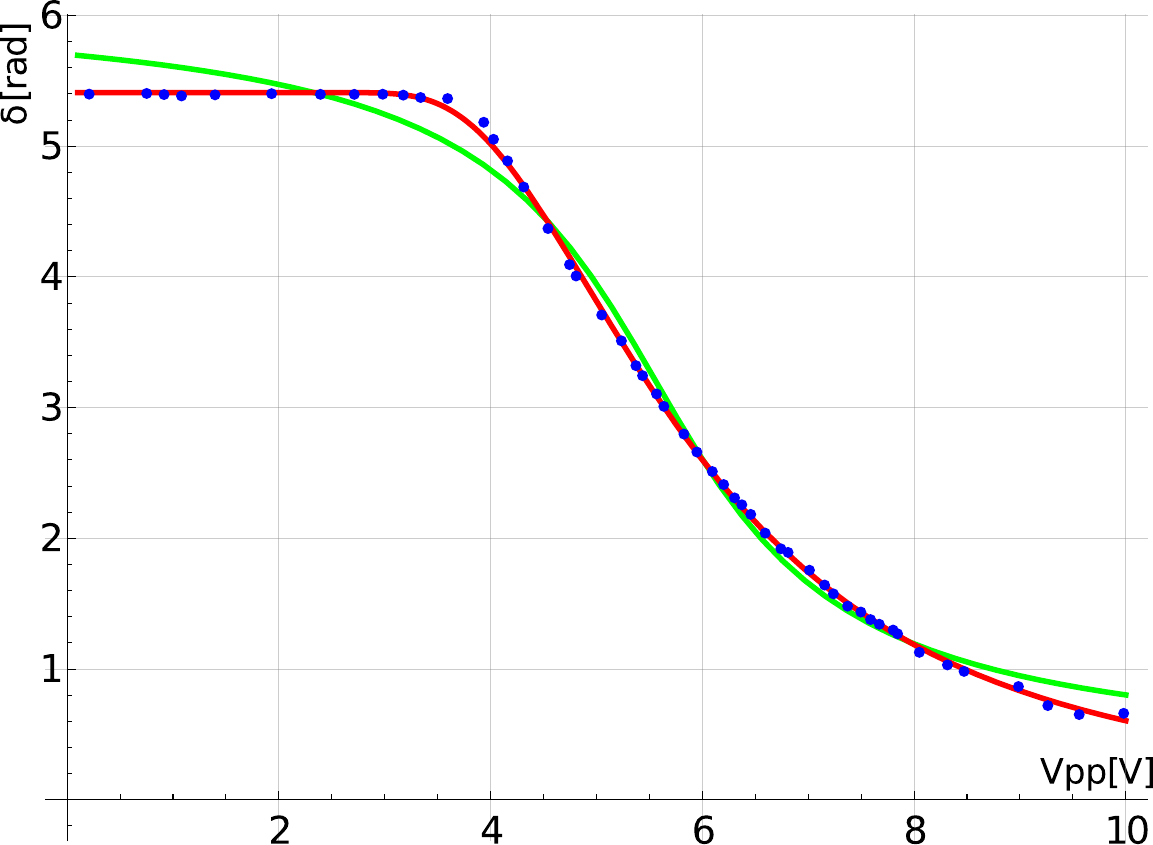}
		}
		\caption {
			The measured phase delay of a single TNLC cell obtained from the display Lumex LCD-S101D14TR as a function of the applied voltage at 810~nm (blue markers). The delay is described using an $\arctan{}$ function (green) or a logistic function (red).
		}
		\label{Fig. logistic}
	\end{center}
\end{figure} 

\begin{figure}[!hbt]
	\begin{center}
		\makebox[\linewidth]{
			\includegraphics[width=0.9\textwidth] {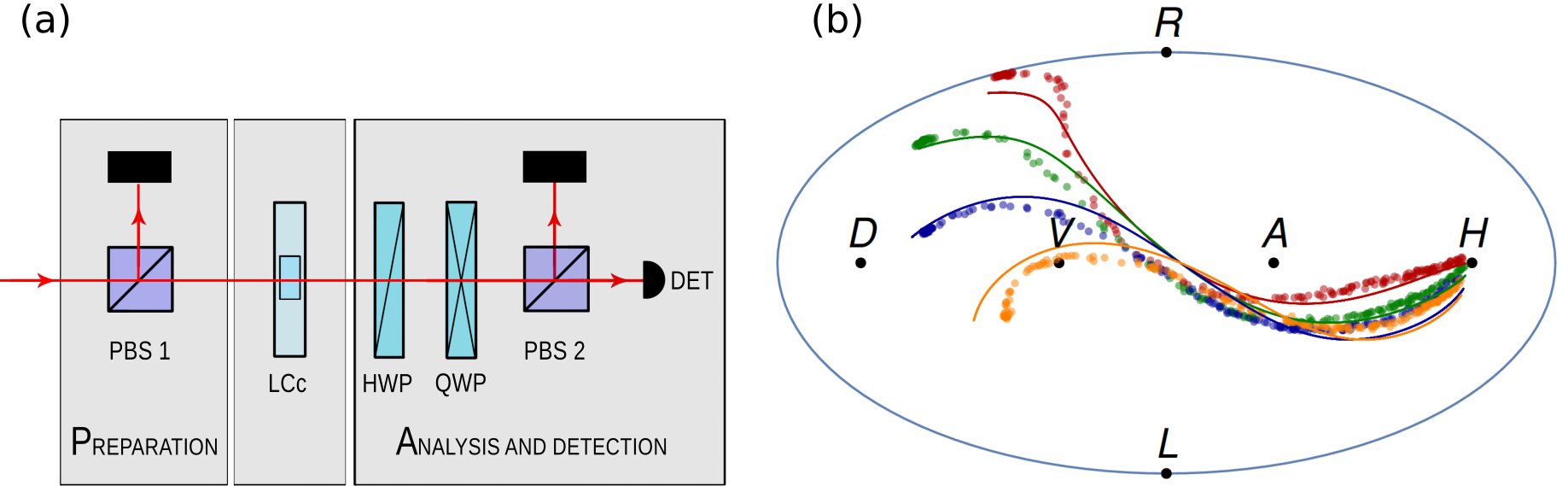}
		}
		\caption {(a) Scheme of the setup used for TNLC cell characterization. The following components are employed: tested TNLC cell (LCc), a polarizing beam splitter (PBS), quarter-wave plate (QWP), half-wave plate (HWP), detector (DET) -- photodiode. The \emph{detection and analysis} block serves as a reference polarimeter. (b) Transformation of the horizontally polarized light by single TNLC cell in Hammer map projection. Dots represent measured states. Solid lines represent fit. Additional display rotation angles 0~deg, 15~deg, 30~deg, and 45~deg are indicated by red, green, blue, and orange colors in this order.}
		\label{Fig. fit}
	\end{center}
\end{figure}

Let us show how a single TNLC cell transforms a polarization state, according to the model, and compare it with experimental data. We prepared state $\ket{H}$ on the input of the tested TNLC cell and changed the control voltage amplitude. For each voltage, we measured the output polarization state using a reference polarimeter. We plot the phase delay introduced by the cell in Fig.~\ref{Fig. logistic} as a function of voltage. The least-square fit with the proposed logistic function in Fig.~\ref{Fig. logistic} clearly describes the measured data better than the approximation with $\arctan$. This measurement was repeated for four known angular positions $\beta$ of the TNLC cell, namely $\beta = 0$~deg, 15~deg, 30~deg, and 45~deg. See Fig.~\ref{Fig. fit}(a) for a scheme of the experimental setup used in this measurement. We performed a least-square fit of the measured data to obtain seven independent parameters of the TNLC model $A=-7.8$, $B=0.076$, $C=176.16$, $D=-2.9$, $E=-2.9$, $\varphi=-5.16$, and $\alpha=-2.6$. 

In Fig.~\ref{Fig. fit}(b), we compare measured data with the theoretical predictions based on fitted parameters. The prepared and expected states are depicted using the Bloch sphere, which is plotted in the Hammer map projection \cite{Snyder1993}. Clearly, a single TNLC cell covers only a curve on the Bloch sphere. To transform state $\ket{H}$ into an arbitrary pure polarization state, one needs to introduce additional independent polarization transformations. This is the reason why the reported TNLC device consists of three TNLC cells controlled with independent voltages. Furthermore, the model deviates from the measured data even though it was fitted to the same data. Parameters of the model vary between individual TNLC cells, which represents a complication in modeling multi-cell devices. The accuracy of the model could be further improved by considering edge effects in the vicinity of aligning layers \cite{Marquez2000,Yamauchi2005} and accounting for other imperfections, such as multiple reflections, depolarization effects, and thickness inhomogeneity. 
However, no model with sufficiently high fidelity with the experimental data covering all possible states has been reported. Also, it is very complicated to make an inverse of such a complex numerical model. In the following section, we introduce a novel calibration method that solves both issues.

\section{Calibration of TNLC device}
\label{sec:devcalib}

We aim for the device able to prepare an arbitrary polarization state with high precision. The calibration of the TNLC device consists of finding control voltages $V_1, V_2$, and $V_3$, one for each TNLC cell, for which the device transforms the pure input state $\ket{H}$ to the pure particular target state $\ket{\psi}$. 

\begin{figure}[!hbt]
	\begin{center}
		\makebox[\linewidth]{
			\includegraphics[width=0.84\textwidth] {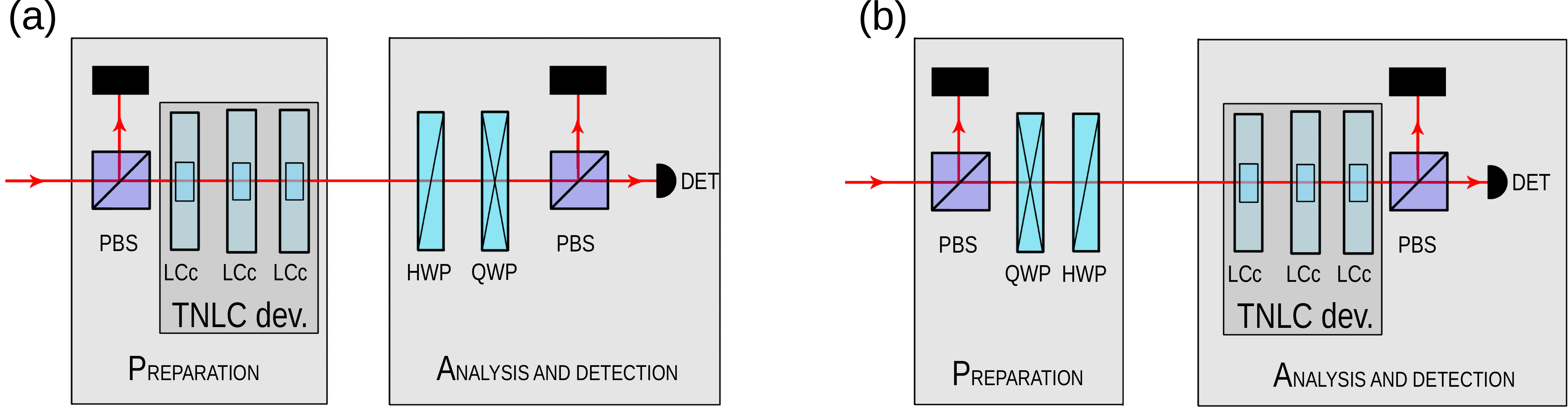}
		}
		\caption {Schemes of the calibration (a) and the polarimetric measurement using the TNLC device (b). The following components are employed: the TNLC device consisting of three TNLC cells (LCc), polarizing beam splitter (PBS), quarter-wave plate (QWP), half-wave plate (HWP), detector (DET) -- either photodiode or single photon detector.}\label{Fig. scheme}
	\end{center}
\end{figure}

The calibration is thus formulated as \emph{maximization of the fidelity} $F = |\bra{\psi}M(V_1, V_2, V_3)\ket{H}|^2$, where $M$ is transformation introduced by the TNLC device, over the three control voltages. The fidelity $|\bra{\psi}M(V_1, V_2, V_3)\ket{H}|^2$ is proportional to the intensity of projection onto the target state, which is directly experimentally measurable. The maximization \emph{does not rely} on the theoretical model, and various methods could be used to solve it. Straightforward methods, like brute-force search or random sampling, are time-consuming as they require many measurements. Therefore we choose a genetic algorithm \cite{Eiben2015} to speed up the calibration.

The initial population is created as three-element random-voltage vectors, and a fitness function is the detected photocurrent. We sequentially apply these voltages settings and directly measure the value of the fitness function. We choose voltages with the largest fitness from this first generation and crossover them to produce ``child'' voltages, which replace the voltages of lesser fitness. As a crossover method, we chose the arithmetic mean of the corresponding voltages. A random mutation in range of $\pm500$~mVpp is also applied to a randomly chosen component of the voltage vector. These steps are repeated until the fitness changes less than one part per thousand, which is roughly an order of magnitude larger than the uncertainty of the intensity measurement. Further iterations would slightly improve the precision at the expense of the longer time needed for the calibration. Although the TNLC theoretical model (\ref{eq: TNLC matrix})--(\ref{eq: Logistic}) is neither perfect nor required, it can be used to speed up the search by providing a rough initial estimate of the voltage values for the first generation of the genetic algorithm.
Even faster calibration might be obtained using black-box optimization methods \cite{Rios2012}. As the last step of the calibration, we generate 120 random voltage vectors around the fittest voltage vector using a normal distribution with a half-width of 0.1~Vpp. We prepare states using these voltages and measured them using a reference polarimeter. Finally, we select the optimum voltages that produce the target state with the highest fidelity.

The optimum voltages slightly vary for the TNLC cells extracted from different TNLC displays, even when selected from a single manufacturing batch. The voltages are also affected by offsets of voltage regulators and digital-to-analog converters incorporated in the electronic drivers, see Sec.~\ref{sec:construction}. Consequently, the calibration has to be performed independently for each TNLC device. A detailed calibration sheet with optimum voltages for various target states and several TNLC devices assembled in our lab is available on GitHub \cite{GitHub}.

The experimental setup for the calibration is depicted in Fig.~\ref{Fig. scheme}(a). An 810-nm continuous-wave laser light passes the first horizontal polarizer (calcite crystal with extinction ratio exceeding $1\;:\;10^4$), and then it passes the TNLC device. The output polarization state is projected onto the desired state $\ket{\psi}$ in a detection block, consisting of half- and quarter-wave plates followed by another horizontal polarizer (another calcite crystal). The photodiode measures the optical intensity, which is proportional to the fitness function. The wave plates are mounted in motorized rotation stages (Newport PR50CC) with the angular speed of 20~deg/s and typical bi-directional repeatability $\pm30$~mdeg. The angular positions of these wave plates control which projection we measure. We also use the waveplates to realize projections to six eigenstates of Pauli matrices, $\ket{H}$, $\ket{V}$, $\ket{D}$, $\ket{A}$, $\ket{R}$, and $\ket{L}$. Section~\ref{sec:polarimter} explains how to use these projections to reconstruct a quantum state. For now, we state that the \emph{analysis and detection} block serves also as a reference polarimeter.

In addition to \emph{preparing} arbitrary pure polarization state, the TNLC device can \emph{project} a state onto an arbitrary pure state. Just by simple reversion of the calibration setup, depicted in Fig.~\ref{Fig. scheme}(b), we could maximize $F = |\bra{H}M(V_1, V_2, V_3)\ket{\psi}|^2$ using the same algorithm.
The presented calibration can be performed for an arbitrary wavelength within visible and near-infrared regions.

\section{Precise preparation of arbitrary polarization}
\label{sec:preparation}
We first test the presented TNLC device calibration on a set of six eigenstates of Pauli operators. We measured the prepared state with the reference polarimeter. The Bloch parameters of the prepared states are shown in Table~\ref{tab.: HVDARL states 3TNLCd}. The prepared states show high purity, $P = 0.999(2)$ on average, and are very close to the ideal target states.
The number in parenthesis represents one standard deviation at the last decimal digit~\cite{STD_NIST2020Jul}. The average fidelity of the observed state to the ideal target states reaches $F = 0.999(1)$ with the corresponding average angular deviation of 1.1(3)~deg. We believe that the very small remaining discrepancy is mainly due to the stopping criteria of the calibration search (one part per thousand) and also inaccuracies of the reference polarimeter, such as retardation error of the waveplates or their misalignment.

We observed perfect stability in time and low sensitivity to temperature changes under common laboratory conditions. The cell's temperature variation within $\pm3^{\circ}$C did not cause any measurable changes in the calibration. Similarly, central wavelength variation within $\pm5$~nm does not introduce any measurable changes.

\begin{table}[!hbt]
	\small
	\centering	
		\makebox[\linewidth]{
			\begin{tabular}{|M{1.6cm}||c|c|c|c|c|c|}
				\hline
				Bloch parameters&    H    & V    & D    & A    & R     & L  \\ \hline \hline
				$u_1$ &    \textbf{0.99941(5)}&    \textbf{-0.99930(3)}&    -0.040(2)&                  -0.0079(6)&              -0.0192(5)&              0.023(2)\\ \hline
				$u_2$ &     0.028(1)&             -0.0355(9)&              \textbf{0.99918(7)}&        \textbf{-0.99723(6)}&    -0.0307(2)&              0.047(3)\\ \hline
				$u_3$ &     0.0199(4)&             0.0112(4)&               -0.0004(7)&                 -0.0186(1)&               \textbf{0.997827(9)}&    \textbf{-0.9938(3)}\\ \hline
			\end{tabular}
		}
	\caption{Bloch parameters of H, V, D, A, R, and L polarization states prepared by the TNLC device and measured by the reference polarimeter.}
	\label{tab.: HVDARL states 3TNLCd}
\end{table}
\begin{figure}[!hbt]
	\begin{center}
		\makebox[\linewidth]{
			\includegraphics[width=0.45\textwidth]{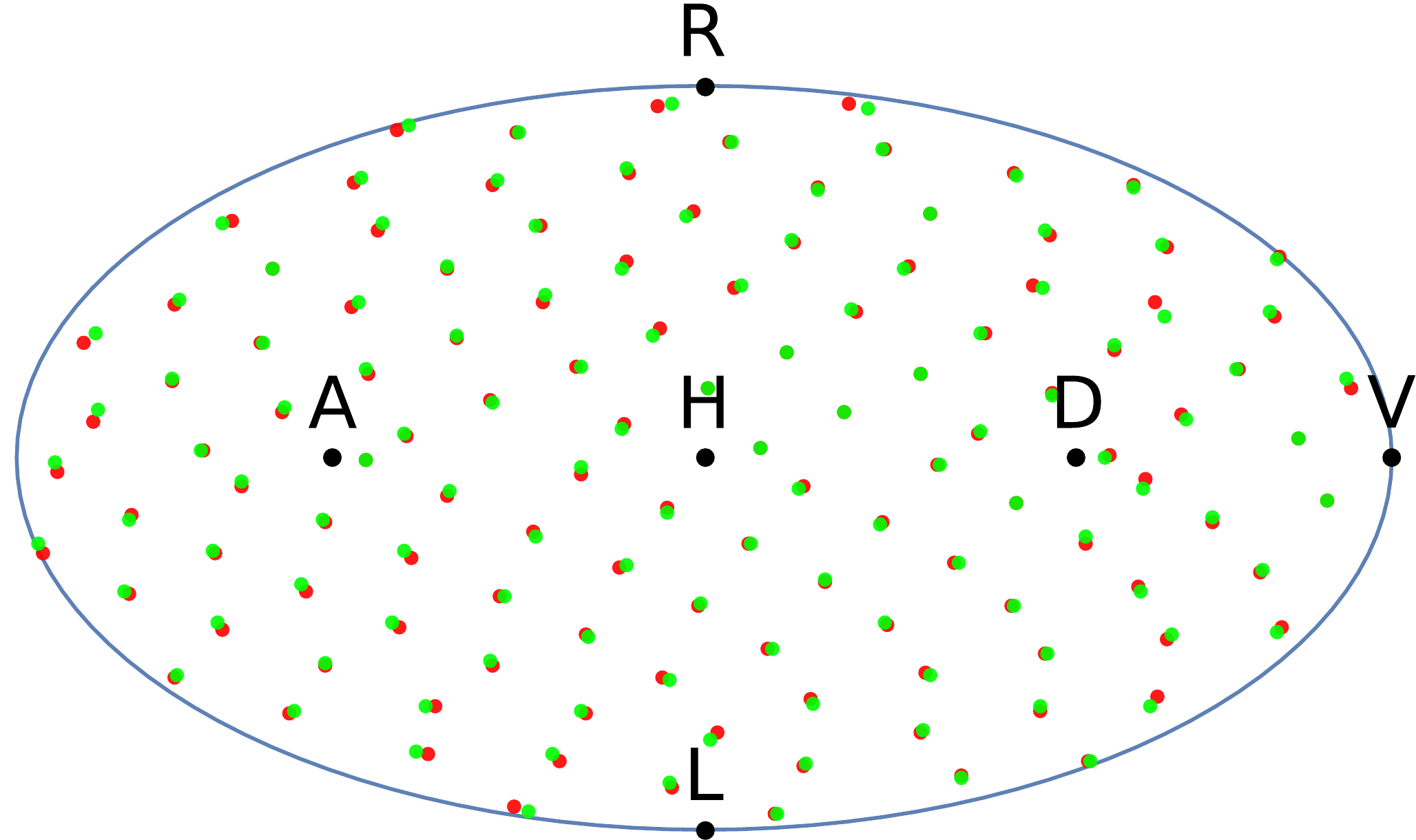}
		}
		\caption{Demonstration of the Bloch sphere coverage using the reported TNLC device. The target states are represented by red points, green markers show the states prepared by the TNLC device. The sphere is visualized in the Hammer map projection.}
		\label{Fig. BlochLCD}
	\end{center}
\end{figure}

We chose the six eigenstates of Pauli operators because they are traditionally used in polarization state tomography, which we will show in Section~\ref{sec:polarimter}. Another important tomographic set of states forms the vertices of a regular tetrahedron inscribed into the Bloch sphere. We searched for the states that maximize the tetrahedron volume and reached the ratio of the measured volume to the volume of the regular tetrahedron $V_\text{meas}/V_\mathrm{ideal} = 0.9987(5)$.

Furthermore, we prepare 120 states, quasi-uniformly distributed over the Bloch sphere. Figure~\ref{Fig. BlochLCD} visually compares the prepared states with the ideal target states. It clearly shows that the presented device can produce an arbitrary pure state with high precision. The prepared states were again almost pure, with average purity $P = 0.999(2)$. The average fidelity of 0.999(1) and the corresponding average angular deviation of 0.5(3)~deg indicate the precise preparation of arbitrary pure state achieved by the presented device.

In addition to the pure polarization states, it is also beneficial to prepare partially mixed states, i.e., states with an arbitrary degree of polarization DoP. Depolarization can be achieved by ensemble averaging of a polarization state in spatial, frequency, or time domain \cite{Peinado2014}. Often, the maximally depolarized state is prepared and then superimposed to a pure state to form a source with continuously tunable DoP. In the case of a voltage-controlled TNLC cell, a specific voltage applied to the TNLC cell performs corresponding unitary operation. Therefore, time multiplexing of various control voltages within the single acquisition time must be used.
We demonstrate the preparation of a maximally mixed state by uniformly cycling through the polarization states H, V, D, A, R, and L during the measurement time. Note that the acquisition is paused during the time needed to switch from one prepared state to another. The resulting prepared state possesses the purity 0.5004(3) and the corresponding DoP 0.03(1).

\begin{table}[hbt] 
	\begin{center}
		\makebox[\linewidth]{\small{
				\begin{tabular}{|ll|rrrrrr|}
					\hline
					\multicolumn{2}{|c|}{\multirow{2}{*}{}}  & \multicolumn{6}{c|}{$\ket{\pi_{\mathrm{fi}}}\bra{\pi_{\mathrm{fi}}}$}   \\
					\multicolumn{2}{|c|}{}  & H  & V  & D  & A  & R  & L  \\
					\hline				
					\multirow{6}{*}{\rotatebox{90}{$\ket{\pi_{\mathrm{in}}}\bra{\pi_{\mathrm{in}}}$}}
					& H &      & 49  & 226 & 590 & 251 & 319 \\
					& V & 168  &     & 182 & 597 & 271 & 329 \\
					& D & 313  & 232 &     & 200 & 203 & 255 \\
					& A & 444  & 403 & 188 &     & 312 & 197 \\
					& R & 566  & 151 & 417 & 481 &     & 141  \\
					& L & 601  & 497 & 299 & 185 & 424 &    \\
					\hline
			\end{tabular}}
		}
		\caption{Transition times in milliseconds between the initial $\ket{\pi_{\mathrm{in}}}\bra{\pi_{\mathrm{in}}}$ and final polarization projection $\ket{\pi_{\mathrm{fi}}}\bra{\pi_{\mathrm{fi}}}$ of the TNLC device.}
		\label{tab.: times}
	\end{center}
\end{table}

The reported TNLC device is made from thick TNLC cells and is not optimized with respect to its speed. However, to present the full characterization of the TNLC device, we measured its time response, for which we used the setup shown in Fig. \ref{Fig. scheme}(b). The wave plates prepare a pure input state which is then projected onto another pure state using the tested device. The corresponding projection intensity is detected with a photodiode. To study the time response, we switch the projection and sample the transition in the detected photocurrent using an analog-to-digital converter with an 18-bit resolution acquiring approximately 1000 samples per second.

\begin{figure}[!h!]
	\centering
	\makebox[\linewidth]{
		\includegraphics[width=0.95\textwidth] {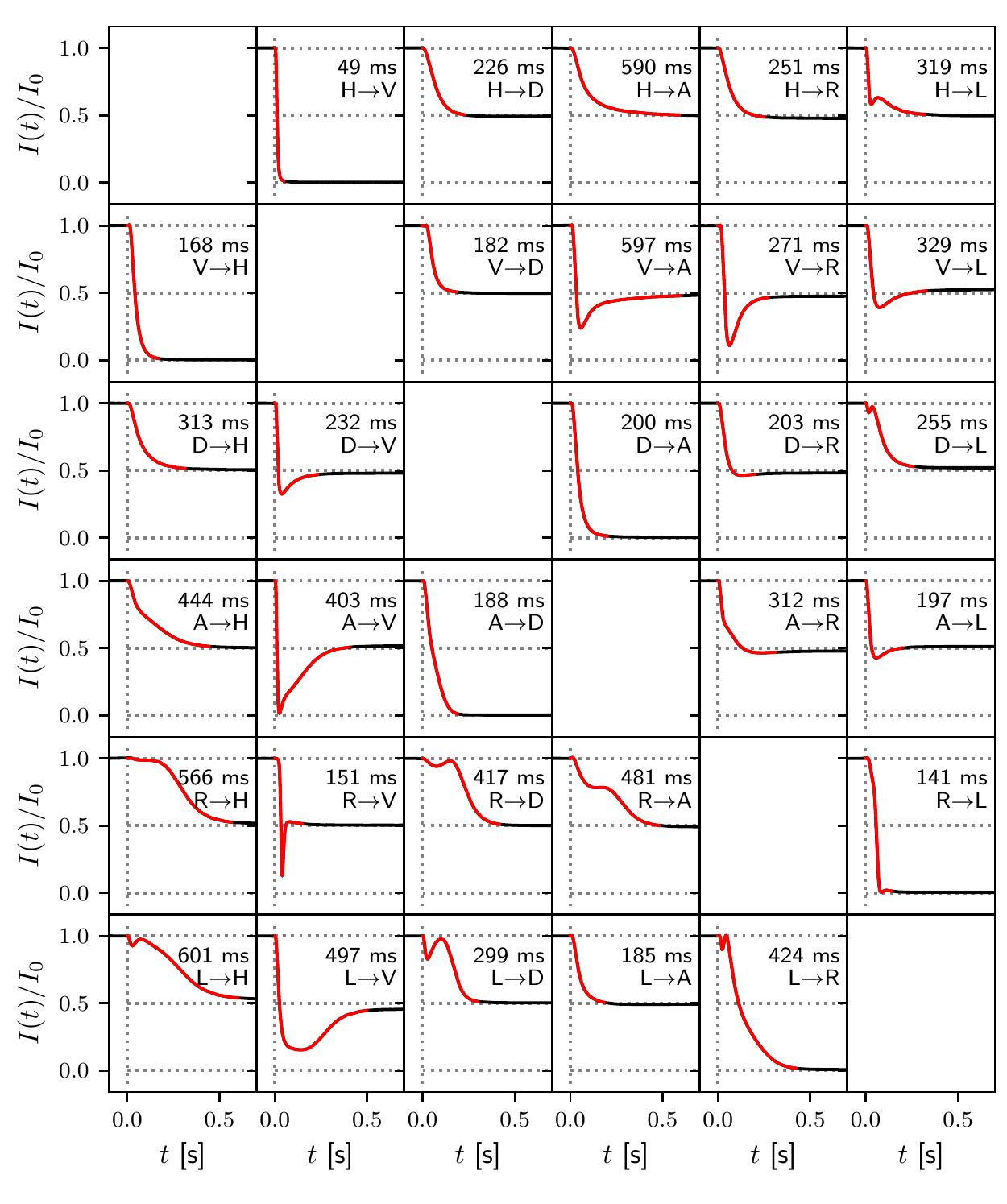}
	}
	\vspace{-8mm}
	\caption{Time traces of transitions between two projections realized by the TNLC device. Initial projections are arranged in rows and the final projections are arranged in columns. For each transition from initial to final projection, there is a time trace of measured intensity. Red color highlights the transition time. The horizontal axes in all plots represent time in seconds while the vertical axes represent normalized intensity.}
	\label{Fig.: times}
\end{figure}

Specifically, we measured the transition times of all the projections onto Pauli operators eigenstates (H, V, D, A, R, L). We prepared a corresponding input state for an initial projection to obtain a maximal optical intensity at the detector. Then we switched the projection. We define the \emph{transition time} as the time from the issued command to the moment when the intensity deviation from its final value stays under $1\%$ of the total intensity ($0\%$ to $99\%$ transition time). We chose this value to ensure a stable operation after the transition. The measured transition times are shown in Table~\ref{tab.: times}, and the time traces of all 30 transitions are shown in Fig.~\ref{Fig.: times}.

The particular transitions, such as H$\rightarrow$V, are rather fast and suitable for rapid polarization switching, but there is a transition exceeding 600~ms, namely L$\rightarrow$H. Specially designed nematic liquid crystal retarders and switches typically specify faster switching, particularly those with a thinner liquid crystal layer and a single optimized polarization transition. Also, their transition times are often specified as $10\%$ to $90\%$ or even $20\%$ to $80\%$, which is not sufficient for precise settling of the induced polarization transformation in polarimetry.

\section{TNLC polarimeter}
\label{sec:polarimter}
The presented device is compact, robust, and capable of precise polarization manipulation with less than one second transition times. Here, we utilize the developed TNLC device to create a polarimeter based on \emph{polarization state tomography}. The polarization state tomography finds its applications mainly when we deal with single-photon signals or measure multi-photon polarization states. In these applications, the classical polarimetry techniques are not sufficient. The presented polarimeter outperforms conventional rotating-waveplates-based polarization tomography in speed, precision, and versatility. 

Let us first briefly review the principle of polarization state tomography. The goal of the polarization state tomography is to unambiguously reconstruct an unknown state $\rho$ from a set of measurements. Let us start with the strong signal and six-state tomography. We assume that the light has intensity $I_0$ and unknown polarization. Recall that any density matrix could be decomposed as a sum of Pauli operators multiplied by Bloch vector components $\rho = (\eye + \sum_{j=1}^{3} u_j \sigma_j )/2$. The projections onto eigenstates of these operators, $\ket{H}$, $\ket{V}$, $\ket{D}$, $\ket{A}$, $\ket{R}$, and $\ket{L}$ contains information about the polarization. Specifically, the measured intensities of these projections are
\begin{eqnarray}
\label{eq:sixstatetomo1.1}
I_{H} = I_{0} \bra{H}\rho\ket{H} = I_{0} (1 + u_1), \\
\label{eq:sixstatetomo1.2}
I_{V} = I_{0} \bra{V}\rho\ket{V} = I_{0} (1 - u_1), \\
\label{eq:sixstatetomo1.3}
I_{D} = I_{0} \bra{D}\rho\ket{D} = I_{0} (1 + u_2), \\
\label{eq:sixstatetomo1.4}
I_{A} = I_{0} \bra{A}\rho\ket{A} = I_{0} (1 - u_2), \\
\label{eq:sixstatetomo1.5}
I_{R} = I_{0} \bra{R}\rho\ket{R} = I_{0} (1 + u_3), \\
\label{eq:sixstatetomo1.6}
I_{L} = I_{0} \bra{L}\rho\ket{L} = I_{0} (1 - u_3).
\end{eqnarray}
These six equations contain three unknown parameters $u_i$ of the polarization state. In the previous section, we showed that our device is capable of realizing these projections. 
One can obtain the parameters $u_i$ from (\ref{eq:sixstatetomo1.1}-\ref{eq:sixstatetomo1.6}) by straightforward inversion,
\begin{eqnarray}
u_1 = \frac{I_{H} - I_{V}}{I_{H} + I_{V}},\\
u_2 = \frac{I_{D} - I_{A}}{I_{D} + I_{A}},\\
u_3 = \frac{I_{R} - I_{L}}{I_{R} + I_{L}}.
\label{eq:sixstatetomo2}
\end{eqnarray}
In the presence of measurement noise, the linear inversion may introduce artifacts. For example, the reconstructed density matrix can have negative eigenvalues, leading to a degree of polarization greater than one. We call such a density matrix \emph{unphysical}. 

To avoid these problems, we use a maximum-likelihood reconstruction method~\cite{Hradil1997, James2001, Jezek2003, Hradil2004}. The method searches for a valid quantum state that would reproduce the measured data with the maximum likelihood. It guarantees that the reconstructed state is always physical. Instead of optical intensities $I_i$ of projection $\bra{\pi_i}\rho\ket{\pi_i}$, we work with relative frequencies $f_i$, which correspond to normalized intensities $f_i = \frac{I_{i}}{\sum I_{i}}$. The logarithm of likelihood of observing relative frequencies $\{ f_i \}$ given the input polarization state $\rho$ is proportional to
\begin{equation}
\ln \mathcal{L}(\rho) = \sum_{i}f_i \ln \bra{\pi_i}\rho\ket{\pi_i}.
\label{eq:likelihood}
\end{equation}
The likelihood quantifies the degree of belief in the hypothesis that for the particular set of observations the polarization is in state $\rho$. We can work with the logarithm of the likelihood with neglected multiplicative factors because if the likelihood is maximized, then any monotonically increasing function of the likelihood is also maximized.

The density matrix $\rho$, which maximizes the likelihood, satisfies the extremal equation \cite{Hradil1997, Jezek2003, Hradil2004}
\begin{equation}
K\rho = \rho,
\label{eq:likelihood2}
\end{equation}
where
\begin{equation}
K = \sum\limits_{i}f_i\frac{\ket{\pi_i}\bra{\pi_i}}{\bra{\pi_i}\rho\ket{\pi_i}}.
\label{eq:likelihood3}
\end{equation}
Equation~(\ref{eq:likelihood2}) leads to an iteration that maximizes the likelihood. The iteration reads 
\begin{equation}
\rho_{i+1} = \frac{K_i \rho_i K_i}{\mathrm{Tr}[K_i \rho_i K_i]}
\label{eq:likelihood4}
\end{equation}
starting from $\rho_0 = \eye/2$ and stopping when the change in one step gets small enough, $\mathrm{Tr}|\rho_{i+1} - \rho_i|^2 < \epsilon$. The derivations of the extremal and iterative equations are covered in Ref.~\cite{Hradil1997, Jezek2003, Hradil2004}. So far, we have used projection measurement operators $\Pi_{i} = \ket{\pi_i}\bra{\pi_i}$, namely $\ket{H}\bra{H}, \dots, \ket{L}\bra{L}$. Let us note that the polarization tomography is not restricted to these particular projections and \emph{any tomographically complete} set of positive measurement operators (POVM) is applicable. The method also applies to single-photon signals, for which the detected intensity $I_i$ is replaced with the recorded number of detection events in a given projection. We provide a Python script for the maximum likelihood reconstruction on GitHub~\cite{GitHub}. Examples of waveplate-based and TNLC-based polarization tomography are also included.

We test the TNLC polarimeter by comparing its readings to the known input polarization states. We prepared the input polarization states using waveplates and measured them using the tested TNLC polarimeter. Namely, we prepared the six eigenstates of Pauli operators. The TNLC polarimeter was calibrated to perform the projections onto Pauli eigenstates as well as the four states required for minimal tomography~\cite{Rehacek2004}. As a reference, we replaced the TNLC polarimeter with a pair of waveplates and performed the standard six-state tomography. 

\begin{figure}[!hbt]
	\begin{center}
		\makebox[\linewidth]{
			\includegraphics[width=0.45\textwidth] {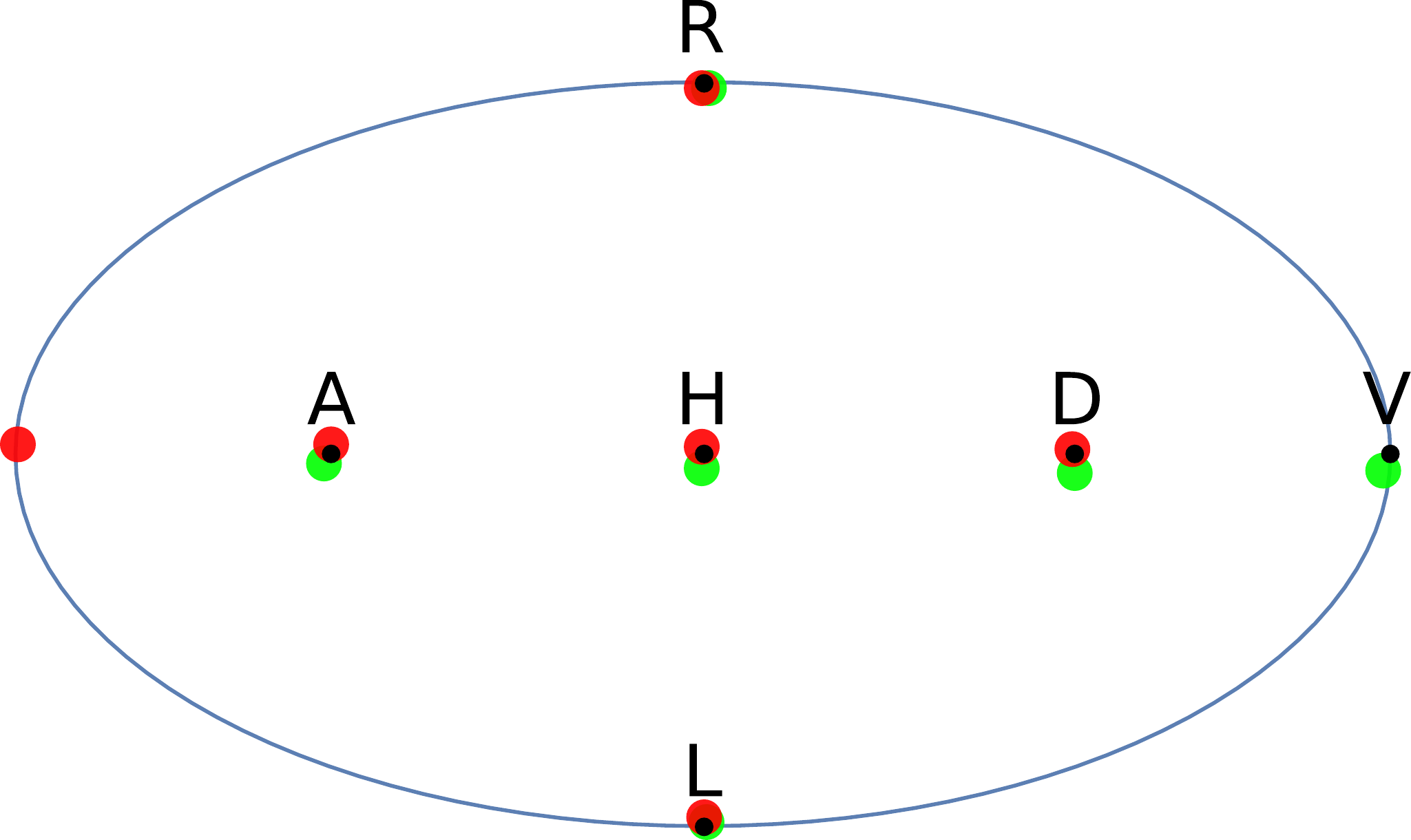}
		}
		\caption{The Bloch sphere with H, V, D, A, R, and L states prepared by waveplates and analyzed by the TNLC device using six-state tomography (red dots) and four-state minimum tomography (green points). Black points show the ideal position of the states.}
		\label{Fig. BlochLCD2}
	\end{center}
\end{figure}

The resulting fidelities and angular deviations averaged over the six tested states are summarized in Table~\ref{tab.: Average results}. The individual sample states analyzed using the presented TNLC device with the six-state tomography and the four-state tomography are visualized in Fig.~\ref{Fig. BlochLCD2}. The six-state tomography outperforms the four-state minimal tomography consistently. The results produced by the TNLC polarimeter agree excellently with the results provided by a reference waveplate-based polarimeter. Note that the used reference polarimeter is not the same polarimeter we used for TNLC device calibration. 

\begin{table} \catcode`\-=12 
	\begin{center}
			\makebox[\linewidth]{
				\begin{tabular}{|M{3cm}||c|c|c|c|c|c|c|c|c|c|c|}
					\hline
					analysis method         & \multicolumn{2}{c|}{TNLC device}         & wave plates \\ \cline{2-4}
					~                    & six-state conf.          & four-states conf.       & six-state conf. \\ \hline \hline 
					fidelity                & 0.9996(4)       & 0.998(2)         & 0.9995(5)    \\ \hline
					angular deviation [deg]  & 1.0(2)        & 1.3(5)           & 1.0(5)    \\ \hline
				\end{tabular}
			}
		\caption{Achieved average results of the fidelity and angular deviation between the ideal and measured sample polarization states prepared by the reference polarimeter. The reported TNLC polarimeter in six-state and four-state configurations, and the reference wave-plate based polarimeter in six-state regime are compared.}
		\label{tab.: Average results}
	\end{center}
\end{table}

Now we show that the presented TNLC polarimeter outperforms the equivalent scheme with waveplates in terms of speed. The ordering of the projections in the tomographic measurement does not affect the result of the measurement but influences its duration \cite{Hosak2018}. Based on the transition time characterization (Table \ref{tab.: times}), we can select the optimum sequence ordering of the six-state protocol, which is H, R, L, A, D, V, and back to H to prepare the TNLC device for another tomography cycle. The transition of the optimum sequence takes 1.17~s. For the minimal tomography~\cite{Rehacek2004}, only four states are required. The transitions times between them are longer than in the six-state scheme. It takes approximately 1.2~s to cycle through these four states. Consequently, we recommend to using six-state tomography with the presented device.

We compare these values with the total time required by our reference waveplate-based polarimeter. The reference polarimeter uses motorized mounts with a typical speed of 20~deg/s (Newport PR50CC), and the optimum sequence reads H, L, A, R, V, D, and back to H. The sequence takes 12.9 s, which is an order of magnitude slower than using the TNLC device. The optimal sequence ordering can significantly speed up the whole measurement process, especially for multi-mode analysis or quantum circuits with more qubits \cite{Hosak2018}. Alternatively, all projections can be measured at the same time \cite{Ling2006,Estevez2016} at the expense of a number of detectors and the complexity of the experimental setup. 

\begin{figure}[!hbt]
	\begin{center}
		\includegraphics[width=0.75\textwidth] {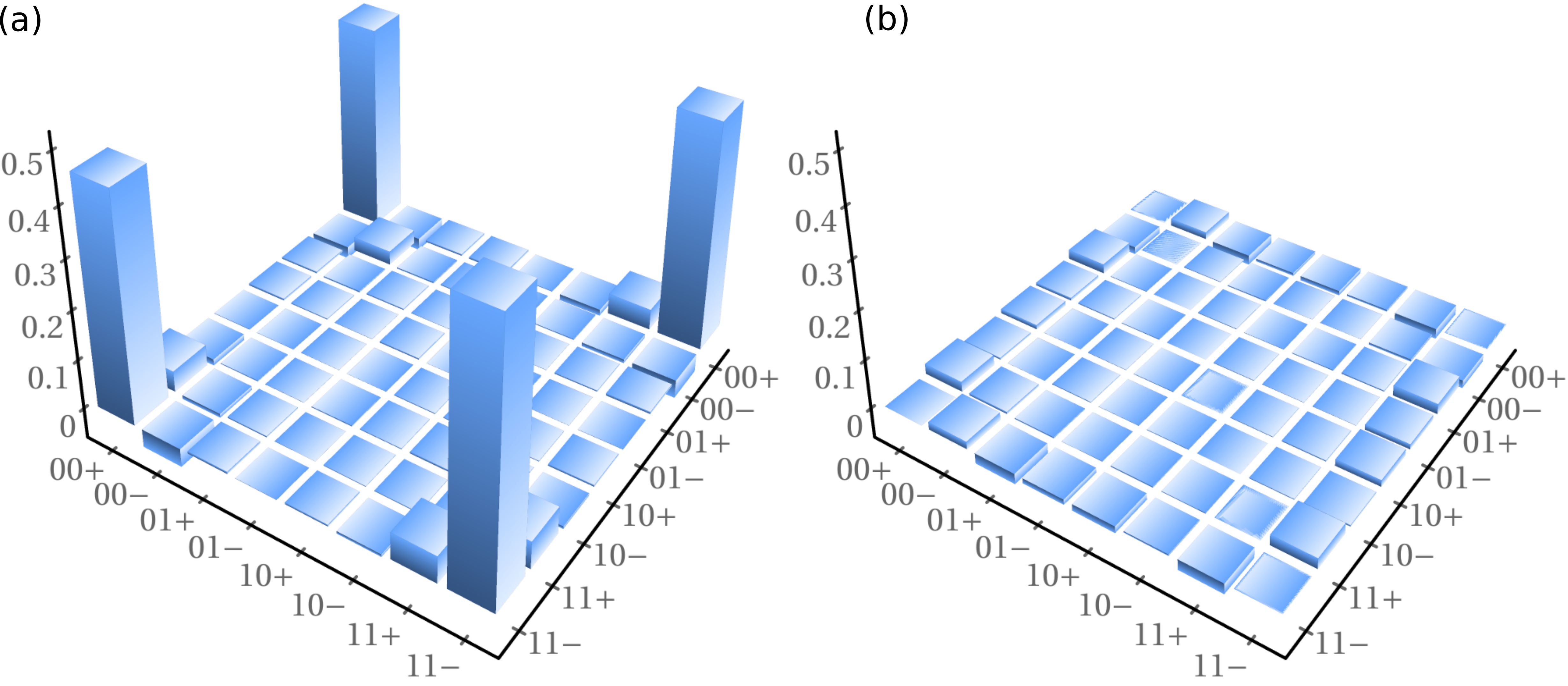}
		\includegraphics[width=0.75\textwidth] {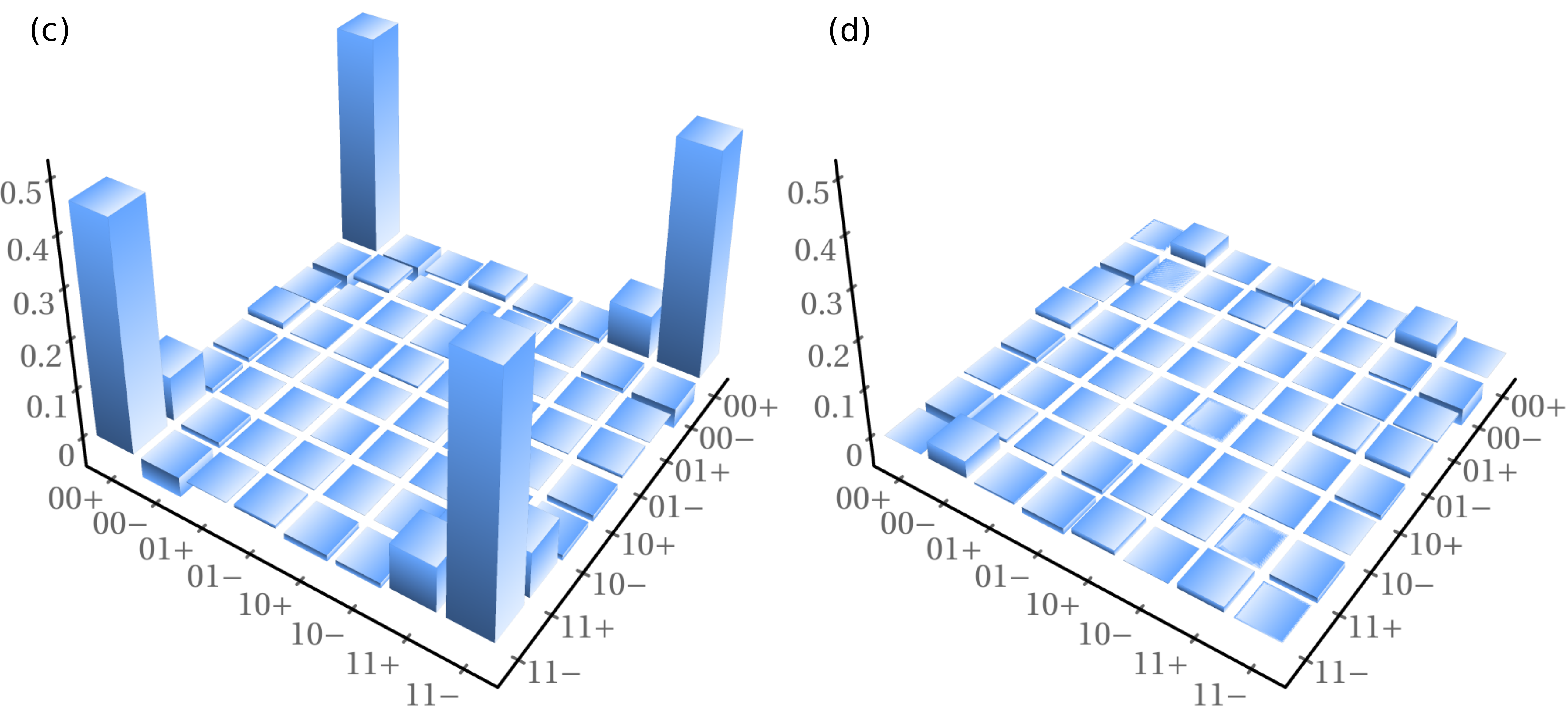}
		\caption {Real (a) and imaginary (b) part of the GHZ state density matrix measured using the TNLC device. Real (c) and imaginary (d) part of the GHZ state density matrix measured using wave plates. Here 0, 1, +, and $-$ denote H, V, D, and A polarizations, respectively.}
		\label{Fig. GHZ}
	\end{center}
\end{figure} 

To show an application of the reported TNLC device at the single-photon level, we used it in quantum tomography of a multi-qubit photonic entangled state. The TNLC device was calibrated in a six-state configuration using the strong laser signal, as described in Sec.~\ref{sec:devcalib}. The TNLC device was then used to perform tomographic projections on the third output qubit of a three-qubit linear optical controlled-Z gate (CCZ gate). The gate is equivalent to the Toffoli gate up to single-qubit Hadamard transform on a target qubit \cite{Nielsen2000,Lanyon2008,Monz2009,Fedorov2011}. The Toffoli gate is a crucial part of many quantum information processing schemes \cite{Nielsen2000}.
The photonic circuit of the gate and its characterization were presented in detail elsewhere \cite{Starek2018}. The gate can produce entangled quantum states such as tripartite Greenberger-Horne-Zeilinger (GHZ) state \cite{Greenberger1989,Pan2000}. We used the circuit to prepare the GHZ state $\left(|HHD\rangle + |VVA\rangle\right)/\sqrt2$ and performed its full tomographic characterization using the TNLC device. The resulting density matrix of the three-qubit state is shown in Fig.~\ref{Fig. GHZ}(a,b). The fidelity of the measured state and the ideal GHZ state is 0.9362.
For comparison, we characterized the state also using a common waveplate-based tomography and obtained the fidelity to the ideal state 0.9360; the resulting density matrix is shown in Fig.~\ref{Fig. GHZ}(c,d).
The respective fidelity of the retrieved GHZ states measured using the TNLC device and the wave-plate polarimeter reaches 0.971. The small resulting discrepancy is caused by the limited repeatability of the experimental setup. When changing the TNLC device for the waveplates, the setup had to be partly reassembled and realigned. Due to the circuit construction, this realignment might lead to small changes in polarization-dependent losses. Furthermore, the interferometric phase in the experimental setup was slowly drifting, and the phase drift might result in small changes between two experimental runs.

\section{Conclusion}
\label{sec:conclusion}

We have presented a polarimetric device assembled from a commercially available TNLC segment display with minimal modifications. We have demonstrated a universal method of the TNLC device calibration utilizing a genetic algorithm. Various polarization states have been prepared using the TNLC device and characterized with a reference polarimeter with the average fidelity exceeding 0.999(1). Particularly, we have demonstrated a highly accurate generation of four states forming the vertices of a regular tetrahedron inscribed into the Bloch sphere~\cite{Rehacek2004}, six eigenstates of Pauli operators (horizontal, vertical, diagonal, anti-diagonal, right-handed and left-handed circular polarizations), and 120 states uniformly distributed on the Bloch Sphere. We also have generated a completely depolarized state with the purity 0.5004(3) and degree of polarization 0.03(1). Using the presented calibration procedure, we can generate an arbitrary state of polarization.

Furthermore, we have employed the reported TNLC device as a polarimeter and demonstrated accurate measurement of polarization of light. Physically correct density matrix of the polarization state is retrieved using the maximum likelihood iterative algorithm. We have tested two tomographic schemes, namely four-projection minimal tomography and overdetermined six-projection scheme based on eigenstates of the Pauli matrices. We have characterized and optimized the precision and speed of the developed polarimeter.
The TNLC device has been successfully used also for polarimetry of optical signals at the single-photon level. Particularly, we have performed the full quantum tomography of the three-qubit Greenberger--Horne--Zeilinger entangled state produced by photonic quantum Toffoli gate. The results agree with those obtained using common wave-plate polarimetry.

To conclude, we have demonstrated the calibration and application in polarimetry of virtually arbitrary \emph{twisted} nematic liquid crystal segment displays. The approach allows highly accurate preparation and measurement of an arbitrary polarization state of light using low-voltage driving and no moving parts. The approach is fully scalable to many optical modes at the single-photon level. It can find applications in a wide range of fields, such as polarization-sensitive biomedical imaging or photonic quantum information processing. Recently, we have developed a deep learning model of TNLC devices, which allows omitting the time-consuming calibration procedure and predicting the TNLC control voltages directly \cite{Vasinka2021Sep}. This novel deep learning approach is particularly beneficial when the polarization needs to be controlled continuously.

The complete technical documentation of the presented TNLC device, including tomography software, is published on GitHub \cite{GitHub}.

\section*{Funding}
The Czech Science Foundation (project 17-26143S);
The Ministry of Education, Youth and Sports of the Czech Republic (project No 8C18002);
European Union's Horizon 2020 (2014--2020) research and innovation framework programme (project HYPER-U-P-S, No 8C18002);
Palack\'y University (projects IGA-PrF-2020-009 and IGA-PrF-2021-006).

\section*{Acknowledgments}
We thank Michal Dudka for developing electronic driver circuitry.
Project HYPER-U-P-S has received funding from the QuantERA ERANET Cofund in Quantum Technologies implemented within the European Union's Horizon 2020 Programme.

\section*{Disclosures}
The authors declare that there are no conflicts of interest related to this article.

\section*{Data availability}
Data underlying the results presented in this paper are not publicly available at this time, but they will be available on the GitHub repository \cite{GitHub} and may be obtained from the authors upon request.

\bibliography{lc_bibliography}

\begin{thebibliography}{10}
\newcommand{\enquote}[1]{``#1''}

\bibitem{Huard1997}
S.~Huard, \emph{Polarization of {L}ight} (Wiley, 1997).

\bibitem{Chen2019}
G.~Chen, Z.-Q. Wen, and C.-W. Qiu, \enquote{{Superoscillation: from physics to
  optical applications},} {\protect\JournalTitle{Light Sci. Appl.}} \textbf{8},
  1--23 (2019).

\bibitem{VanEeckhout2019}
A.~Van~Eeckhout, E.~Garcia-Caurel, T.~Garnatje, M.~Durfort, J.~C. Escalera,
  J.~Vidal, J.~J. Gil, J.~Campos, and A.~Lizana, \enquote{{Depolarizing metrics
  for plant samples imaging},} {\protect\JournalTitle{PLoS One}} \textbf{14},
  e0213909 (2019).

\bibitem{He2021Sep}
C.~He, H.~He, J.~Chang, B.~Chen, H.~Ma, and M.~J. Booth, \enquote{{Polarisation
  optics for biomedical and clinical applications: a review - Light: Science
  {\&} Applications},} {\protect\JournalTitle{Light Sci. Appl.}} \textbf{10},
  1--20 (2021).

\bibitem{Zhanghao2019}
K.~Zhanghao, X.~Chen, W.~Liu, M.~Li, Y.~Liu, Y.~Wang, S.~Luo, X.~Wang, C.~Shan,
  H.~Xie, J.~Gao, X.~Chen, D.~Jin, X.~Li, Y.~Zhang, Q.~Dai, and P.~Xi,
  \enquote{{Super-resolution imaging of fluorescent dipoles via polarized
  structured illumination microscopy},} {\protect\JournalTitle{Nat. Commun.}}
  \textbf{10}, 1--10 (2019).

\bibitem{Chen2016}
Z.-Y. Chen, L.-S. Yan, Y.~Pan, L.~Jiang, A.-L. Yi, W.~Pan, and B.~Luo,
  \enquote{{Use of polarization freedom beyond polarization-division
  multiplexing to support high-speed and spectral-efficient data
  transmission},} {\protect\JournalTitle{Light Sci. Appl.}} \textbf{6}, e16207
  (2017).

\bibitem{Flamini2018}
F.~Flamini, N.~Spagnolo, and F.~Sciarrino, \enquote{{Photonic quantum
  information processing: a review},} {\protect\JournalTitle{Rep. Prog. Phys.}}
  \textbf{82}, 016001 (2018).

\bibitem{Slussarenko2019}
S.~Slussarenko and G.~J. Pryde, \enquote{{Photonic quantum information
  processing: A concise review},} {\protect\JournalTitle{Appl. Phys. Rev.}}
  \textbf{6}, 041303 (2019).

\bibitem{Spagnolo2008}
N.~Spagnolo, C.~Vitelli, S.~Giacomini, F.~Sciarrino, and F.~D. Martini,
  \enquote{{Polarization preserving ultra fast optical shutter for quantum
  information processing},} {\protect\JournalTitle{Opt. Express}} \textbf{16},
  17609--17615 (2008).

\bibitem{He2017}
Y.~He, X.~Ding, Z.-E. Su, H.-L. Huang, J.~Qin, C.~Wang, S.~Unsleber, C.~Chen,
  H.~Wang, Y.-M. He, X.-L. Wang, W.-J. Zhang, S.-J. Chen, C.~Schneider,
  M.~Kamp, L.-X. You, Z.~Wang, S.~H{\ifmmode\ddot{o}\else\"{o}\fi}fling, C.-Y.
  Lu, and J.-W. Pan, \enquote{{Time-Bin-Encoded Boson Sampling with a
  Single-Photon Device},} {\protect\JournalTitle{Phys. Rev. Lett.}}
  \textbf{118}, 190501 (2017).

\bibitem{Takeda2019}
S.~Takeda, K.~Takase, and A.~Furusawa, \enquote{{On-demand photonic
  entanglement synthesizer},} {\protect\JournalTitle{Sci. Adv.}} \textbf{5},
  eaaw4530 (2019).

\bibitem{Tiedau2019}
J.~Tiedau, E.~Meyer-Scott, T.~Nitsche, S.~Barkhofen, T.~J. Bartley, and
  C.~Silberhorn, \enquote{{A high dynamic range optical detector for measuring
  single photons and bright light},} {\protect\JournalTitle{Opt. Express}}
  \textbf{27}, 1--15 (2019).

\bibitem{Altepeter2011}
J.~B. Altepeter, N.~N. Oza, M.~Medi{\ifmmode\acute{c}\else\'{c}\fi}, E.~R.
  Jeffrey, and P.~Kumar, \enquote{{Entangled photon polarimetry},}
  {\protect\JournalTitle{Opt. Express}} \textbf{19}, 26011--26016 (2011).

\bibitem{Bueno2000}
J.~M. Bueno, \enquote{{Polarimetry using liquid-crystal variable retarders:
  theory and calibration},} {\protect\JournalTitle{J. Opt. A: Pure Appl. Opt.}}
  \textbf{2}, 216--222 (2000).

\bibitem{DeMartino2003}
A.~De~Martino, Y.-K. Kim, E.~Garcia-Caurel, B.~Laude, and
  B.~Dr{\ifmmode\acute{e}\else\'{e}\fi}villon, \enquote{{Optimized Mueller
  polarimeter with liquid crystals},} {\protect\JournalTitle{Opt. Lett.}}
  \textbf{28}, 616--618 (2003).

\bibitem{Peinado2010}
A.~Peinado, A.~Lizana, J.~Vidal, C.~Iemmi, and J.~Campos,
  \enquote{{Optimization and performance criteria of a Stokes polarimeter based
  on two variable retarders},} {\protect\JournalTitle{Opt. Express}}
  \textbf{18}, 9815--9830 (2010).

\bibitem{Peinado2011}
A.~Peinado, A.~Lizana, J.~Vidal, C.~Iemmi, and J.~Campos, \enquote{{Optimized
  Stokes polarimeters based on a single twisted nematic liquid-crystal device
  for the minimization of noise propagation},} {\protect\JournalTitle{Appl.
  Opt.}} \textbf{50}, 5437--5445 (2011).

\bibitem{Aharon2009}
O.~Aharon and I.~Abdulhalim, \enquote{{Liquid crystal Lyot tunable filter with
  extended free spectral range},} {\protect\JournalTitle{Opt. Express}}
  \textbf{17}, 11426--11433 (2009).

\bibitem{August2013}
Y.~August and A.~Stern, \enquote{{Compressive sensing spectrometry based on
  liquid crystal devices},} {\protect\JournalTitle{Opt. Lett.}} \textbf{38},
  4996--4999 (2013).

\bibitem{Zhuang1999}
Z.~Zhuang, S.-W. Suh, and J.~S. Patel, \enquote{{Polarization controller using
  nematic liquid crystals},} {\protect\JournalTitle{Opt. Lett.}} \textbf{24},
  694--696 (1999).

\bibitem{Moreno2007}
I.~Moreno, J.~L. Mart{\ifmmode\acute{\imath}\else\'{\i}\fi}nez, and J.~A.
  Davis, \enquote{{Two-dimensional polarization rotator using a twisted-nematic
  liquid-crystal display},} {\protect\JournalTitle{Appl. Opt.}} \textbf{46},
  881--887 (2007).

\bibitem{Safrani2009}
A.~Safrani and I.~Abdulhalim, \enquote{{Liquid-crystal polarization rotator and
  a tunable polarizer},} {\protect\JournalTitle{Opt. Lett.}} \textbf{34},
  1801--1803 (2009).

\bibitem{Peinado2014}
A.~Peinado, A.~Lizana, and J.~Campos, \enquote{{Use of ferroelectric liquid
  crystal panels to control state and degree of polarization in light beams},}
  {\protect\JournalTitle{Opt. Lett.}} \textbf{39}, 659--662 (2014).

\bibitem{Sciarrino2017}
A.~S. Rab, E.~Polino, Z.-X. Man, N.~B. An, Y.-J. Xia, N.~Spagnolo, R.~L.
  Franco, and F.~Sciarrino, \enquote{{Entanglement of photons in their dual
  wave-particle nature},} {\protect\JournalTitle{Nat. Commun.}} \textbf{8},
  1--7 (2017).

\bibitem{Sciarrino2018}
A.~Lumino, E.~Polino, A.~S. Rab, G.~Milani, N.~Spagnolo, N.~Wiebe, and
  F.~Sciarrino, \enquote{{Experimental Phase Estimation Enhanced by Machine
  Learning},} {\protect\JournalTitle{Phys. Rev. Appl.}} \textbf{10}, 044033
  (2018).

\bibitem{Lohrmann2019}
A.~Lohrmann, C.~Perumgatt, and A.~Ling, \enquote{{Manipulation and measurement
  of quantum states with liquid crystal devices},} {\protect\JournalTitle{Opt.
  Express}} \textbf{27}, 13765--13772 (2019).

\bibitem{Wang2008}
X.-j. Wang, Z.-d. Huang, J.~Feng, X.-f. Chen, X.~Liang, and Y.-q. Lu,
  \enquote{{Liquid crystal modulator with ultra-wide dynamic range and
  adjustable driving voltage},} {\protect\JournalTitle{Opt. Express}}
  \textbf{16}, 13168--13174 (2008).

\bibitem{Zhu2013}
G.~Zhu, B.~yan Wei, L.~yu~Shi, X.~wen Lin, W.~Hu, Z.~di~Huang, and Y.~qing Lu,
  \enquote{{A fast response variable optical attenuator based on blue phase
  liquid crystal},} {\protect\JournalTitle{Opt. Express}} \textbf{21},
  5332--5337 (2013).

\bibitem{Perumangatt2021}
C.~Perumangatt, T.~Vergoossen, A.~Lohrmann, S.~Sivasankaran, A.~Reezwana,
  A.~Anwar, S.~Sachidananda, T.~Islam, and A.~Ling, \enquote{Realizing quantum
  nodes in space for cost-effective, global quantum communication: in-orbit
  results and next steps,} in \emph{Quantum Computing, Communication, and
  Simulation,}  P.~R. Hemmer and A.~L. Migdall, eds. ({SPIE}, 2021).

\bibitem{Rehacek2004}
J.~{\v{R}}eh{\'{a}}{\v{c}}ek, B.-G. Englert, and D.~Kaszlikowski,
  \enquote{{Minimal qubit tomography},} {\protect\JournalTitle{Phys. Rev. A}}
  \textbf{70}, 052321 (2004).

\bibitem{Ling2006}
A.~Ling, K.~P. Soh, A.~Lamas-Linares, and C.~Kurtsiefer, \enquote{{Experimental
  polarization state tomography using optimal polarimeters},}
  {\protect\JournalTitle{Phys. Rev. A}} \textbf{74}, 022309 (2006).

\bibitem{deBurgh2008}
M.~D. de~Burgh, N.~K. Langford, A.~C. Doherty, and A.~Gilchrist,
  \enquote{{Choice of measurement sets in qubit tomography},}
  {\protect\JournalTitle{Phys. Rev. A}} \textbf{78}, 052122 (2008).

\bibitem{Ling2008}
A.~Ling, A.~Lamas-Linares, and C.~Kurtsiefer, \enquote{{Accuracy of minimal and
  optimal qubit tomography for finite-length experiments},} arXiv:0807.0991
  (2008).

\bibitem{Bogdanov2010}
{\relax Yu}.~I. Bogdanov, G.~Brida, M.~Genovese, S.~P. Kulik, E.~V. Moreva, and
  A.~P. Shurupov, \enquote{{Statistical Estimation of the Efficiency of Quantum
  State Tomography Protocols},} {\protect\JournalTitle{Phys. Rev. Lett.}}
  \textbf{105}, 010404 (2010).

\bibitem{Bogdanov2011}
{\relax Yu}.~I. Bogdanov, G.~Brida, I.~D. Bukeev, M.~Genovese, K.~S. Kravtsov,
  S.~P. Kulik, E.~V. Moreva, A.~A. Soloviev, and A.~P. Shurupov,
  \enquote{{Statistical estimation of the quality of quantum-tomography
  protocols},} {\protect\JournalTitle{Phys. Rev. A}} \textbf{84}, 042108
  (2011).

\bibitem{Koutny2016}
D.~Koutn{\ifmmode\acute{y}\else\'{y}\fi}, Y.~S. Teo, Z.~Hradil, and
  J.~{\ifmmode\check{R}\else\v{R}\fi}eh{\ifmmode\acute{a}\else\'{a}\fi}{\ifmmode\check{c}\else\v{c}\fi}ek,
  \enquote{{Fast universal performance certification of measurement schemes for
  quantum tomography},} {\protect\JournalTitle{Phys. Rev. A}} \textbf{94},
  022113 (2016).

\bibitem{Marquez2000}
A.~Marquez, J.~Campos, M.~J. Yzuel, I.~S. Moreno, J.~A. Davis, C.~C. Iemmi,
  A.~Moreno, and A.~Robert, \enquote{{Characterization of edge effects in
  twisted nematic liquid crystal displays},} {\protect\JournalTitle{Opt. Eng.}}
  \textbf{39}, 3301--3307 (2000).

\bibitem{Yamauchi2005}
M.~Yamauchi, \enquote{{Jones-matrix models for twisted-nematic liquid-crystal
  devices},} {\protect\JournalTitle{Appl. Opt.}} \textbf{44}, 4484--4493
  (2005).

\bibitem{GitHub}
M.~Bielak, \enquote{Github repository,}
  \url{https://github.com/BielakM/polarimeter} (2020).

\bibitem{Lanyon2008}
B.~P. Lanyon, M.~Barbieri, M.~P. Almeida, T.~Jennewein, T.~C. Ralph, K.~J.
  Resch, G.~J. Pryde, J.~L. O{'}Brien, A.~Gilchrist, and A.~G. White,
  \enquote{{Simplifying quantum logic using higher-dimensional Hilbert
  spaces},} {\protect\JournalTitle{Nat. Phys.}} \textbf{5}, 134--140 (2009).

\bibitem{Micuda2013}
M.~Mi{\v{c}}uda, M.~Sedl{\'{a}}k, I.~Straka, M.~Mikov{\'{a}}, M.~Du{\v{s}}ek,
  M.~Je{\v{z}}ek, and J.~Fiur{\'{a}}{\v{s}}ek, \enquote{{Efficient Experimental
  Estimation of Fidelity of Linear Optical Quantum Toffoli Gate},}
  {\protect\JournalTitle{Phys. Rev. Lett.}} \textbf{111}, 160407 (2013).

\bibitem{Xiao2017}
L.~Xiao, X.~Zhan, Z.~H. Bian, K.~K. Wang, X.~Zhang, X.~P. Wang, J.~Li,
  K.~Mochizuki, D.~Kim, N.~Kawakami, W.~Yi, H.~Obuse, B.~C. Sanders, and
  P.~Xue, \enquote{{Observation of topological edge states in
  parity{\textendash}time-symmetric quantum walks},}
  {\protect\JournalTitle{Nat. Phys.}} \textbf{13}, 1117--1123 (2017).

\bibitem{YarivYeh1984}
A.~Yariv and A.~P. Yeh, \emph{Optical waves in crystals: propagation and
  control of laser radiation} (Wiley, 1984).

\bibitem{Davis1998}
J.~A. Davis, I.~Moreno, and P.~Tsai, \enquote{{Polarization eigenstates for
  twisted-nematic liquid-crystal displays},} {\protect\JournalTitle{Appl.
  Opt.}} \textbf{37}, 937--945 (1998).

\bibitem{Snyder1993}
J.~P.~P. Snyder, \emph{Flattening the {E}arth: {T}wo {T}housand {Y}ears of
  {M}ap {P}rojections} (University of Chicago Press, 1993).

\bibitem{Eiben2015}
A.~E. Eiben and J.~E. Smith, \emph{{Introduction to Evolutionary Computing}}
  (Springer, 2015).

\bibitem{Rios2012}
L.~M. Rios and N.~V. Sahinidis, \enquote{{Derivative-free optimization: a
  review of algorithms and comparison of software implementations},}
  {\protect\JournalTitle{J. Global Optim.}} \textbf{56}, 1247--1293 (2013).

\bibitem{STD_NIST2020Jul}
\enquote{{Standard Uncertainty and Relative Standard Uncertainty},}  (2020).
  [Online; accessed 2. Jul. 2020].

\bibitem{Hradil1997}
Z.~Hradil, \enquote{{Quantum-state estimation},} {\protect\JournalTitle{Phys.
  Rev. A}} \textbf{55}, R1561--R1564(R) (1997).

\bibitem{James2001}
D.~F.~V. James, P.~G. Kwiat, W.~J. Munro, and A.~G. White,
  \enquote{{Measurement of qubits},} {\protect\JournalTitle{Phys. Rev. A}}
  \textbf{64}, 052312 (2001).

\bibitem{Jezek2003}
M.~Je{\v{z}}ek, J.~Fiur{\'{a}}{\v{s}}ek, and Z.~Hradil, \enquote{{Quantum
  inference of states and processes},} {\protect\JournalTitle{Phys. Rev. A}}
  \textbf{68}, 012305 (2003).

\bibitem{Hradil2004}
Z.~Hradil, J.~{\v{R}}eh{\'{a}}{\v{c}}ek, J.~Fiur{\'{a}}{\v{s}}ek, and
  M.~Je{\v{z}}ek, \enquote{{3 Maximum-Likelihood Methodsin Quantum Mechanics},}
  in \emph{{Quantum State Estimation},}  (Springer, 2004), pp. 59--112.

\bibitem{Hosak2018}
R.~Ho{\v{s}}{\'{a}}k, R.~St{\'{a}}rek, and M.~Je{\v{z}}ek, \enquote{{Optimal
  reordering of measurements for photonic quantum tomography},}
  {\protect\JournalTitle{Opt. Express}} \textbf{26}, 32878--32887 (2018).

\bibitem{Estevez2016}
I.~Est{\'{e}}vez, V.~Sopo, A.~Lizana, A.~Turpin, and J.~Campos,
  \enquote{{Complete snapshot Stokes polarimeter based on a single biaxial
  crystal},} {\protect\JournalTitle{Opt. Lett.}} \textbf{41}, 4566--4569
  (2016).

\bibitem{Nielsen2000}
M.~A. Nielsen and I.~L. Chuang, \emph{Quantum {C}omputation and {Q}uantum
  {I}nformation} (Cambridge University Press, 2000).

\bibitem{Monz2009}
T.~Monz, K.~Kim, W.~H\"{a}nsel, M.~Riebe, A.~S. Villar, P.~Schindler,
  M.~Chwalla, M.~Hennrich, and R.~Blatt, \enquote{{Realization of the Quantum
  Toffoli Gate with Trapped Ions},} {\protect\JournalTitle{Phys. Rev. Lett.}}
  \textbf{102}, 040501 (2009).

\bibitem{Fedorov2011}
A.~Fedorov, L.~Steffen, M.~Baur, M.~P. da~Silva, and A.~Wallraff,
  \enquote{{Implementation of a Toffoli gate with superconducting circuits},}
  {\protect\JournalTitle{Nature}} \textbf{481}, 170--172 (2012).

\bibitem{Starek2018}
R.~St{\'{a}}rek, M.~Mi{\v{c}}uda, M.~Mikov{\'{a}}, I.~Straka, M.~Du{\v{s}}ek,
  P.~Marek, M.~Je{\v{z}}ek, R.~Filip, and J.~Fiur{\'{a}}{\v{s}}ek,
  \enquote{{Nondestructive detector for exchange symmetry of photonic qubits},}
  {\protect\JournalTitle{npj Quantum Inf.}} \textbf{4}, 1--7 (2018).

\bibitem{Greenberger1989}
D.~M. Greenberger, M.~A. Horne, and A.~Zeilinger, \enquote{{Going Beyond
  Bell{'}s Theorem},} in \emph{{Bell{'}s Theorem, Quantum Theory and
  Conceptions of the Universe},}  (Springer, 1989), pp. 69--72.

\bibitem{Pan2000}
J.-W. Pan, D.~Bouwmeester, M.~Daniell, H.~Weinfurter, and A.~Zeilinger,
  \enquote{{Experimental test of quantum nonlocality in three-photon
  Greenberger{\textendash}Horne{\textendash}Zeilinger entanglement},}
  {\protect\JournalTitle{Nature}} \textbf{403}, 515--519 (2000).

\bibitem{Vasinka2021Sep}
D.~Va{\ifmmode\check{s}\else\v{s}\fi}inka, M.~Bielak, M.~Neset, and
  M.~Je{\ifmmode\check{z}\else\v{z}\fi}ek, \enquote{{Deep learning of
  polarization transfer in liquid crystals with application to quantum state
  preparation},} {\protect\JournalTitle{arXiv:2109.12436}}  (2021).

\end{thebibliography}

\end{document}